\documentclass[prl,showpacs,showkeys,nofootinbib,twocolumn,floatfix,superscriptaddress]{revtex4}

\usepackage{graphicx}  %
\usepackage{amsmath}
\usepackage{bm}  %
\usepackage{dsfont}
\usepackage{ulem}
\usepackage{braket}

\newcommand*{\intsum}{\mathop{\ooalign{\raise0.2pt\hbox{$\int$}%
            \cr\lower0.3pt\hbox{$\Sigma$}}}}

\newcommand{\bea}{\begin{eqnarray}}
\newcommand{\eea}{\end{eqnarray}}
\newcommand{\beq}{\begin{equation}}
\newcommand{\eeq}{\end{equation}}
\newcommand{\bqa}{\begin{eqnarray}}
\newcommand{\eqa}{\end{eqnarray}}

\def\mqo2{{\!\!\!}}

\begin{document}

\title{Efimov Physics around the neutron rich $^{60}$Ca isotope}
\author{G.~Hagen}
\affiliation{Physics Division, Oak Ridge National Laboratory,
Oak Ridge, TN 37831, USA}
\affiliation{Department of Physics and Astronomy, University of
Tennessee, Knoxville, TN 37996, USA}
\author{P. Hagen}
\author{H.-W. Hammer}
\affiliation{Helmholtz-Institut f\"ur Strahlen- und Kernphysik (Theorie)
and Bethe Center for Theoretical Physics, Universit\"at Bonn, 53115 Bonn, 
Germany\\}
\author{L. Platter}
\affiliation{Argonne National Laboratory, Physics Division, Argonne, IL 60439,
USA}
\affiliation{Department of Fundamental Physics, Chalmers University of Technology, 
SE-412 96 Gothenburg, Sweden\\}
\date{\today}

\begin{abstract}
  We calculate the neutron-$^{60}$Ca S-wave scattering phase shifts
  using state of the art coupled-cluster theory combined with modern
  ab initio interactions derived from chiral effective theory. Effects
  of three-nucleon forces are included schematically as density
  dependent nucleon-nucleon interactions. This information is combined
  with halo effective field theory in order to investigate the
  $^{60}$Ca-neutron-neutron system. We predict correlations between
  different three-body observables and the two-neutron separation
  energy of $^{62}$Ca. This provides evidence of Efimov physics along
  the Calcium isotope chain. Experimental key observables that
  facilitate a test of our findings are discussed.
\end{abstract}

\smallskip
\maketitle

\paragraph{Introduction -}
The emergence of new degrees of freedom is one of the most important
aspects of the physics along the neutron ($n$) drip line. For example,
halo nuclei are characterized by a tightly bound core ($c$) and
weakly-bound valence
nucleons~\cite{Zhukov-93,Jensen-04,Frederico:2012xh,Riisager-12} and
thus display a reduction in the effective degrees of freedom. They are
usually identified by an extremely large matter radius or a sudden
decrease in the one- or two-nucleon separation energy along an isotope
chain. The features of these halos are universal if the small
separation energy of the valence nucleons is associated with a large
S-wave scattering length. These phenomena are then independent of the
details of the microscopic interaction and occur in a large class of
systems in atomic, nuclear and particle
physics~\cite{Braaten-05,Platter:2009gz}.  For a three-body system
(e.g.~core-nucleon-nucleon) interacting through a large S-wave
scattering length, Vitaly Efimov showed that the system will display
discrete scale invariance~\cite{Efimov-70}. This discrete scale
invariance is exact in the limit of zero-range interactions and
infinite scattering lengths. For fixed finite values of the scattering
length and range, it is approximate.  The hallmark feature of this
so-called Efimov effect is a tower of bound states. The ratio of the
binding energies of successive states is characterized by a discrete
scaling factor. This scaling factor is approximately 515 in the case
of identical bosons. Systems whose particles have different masses
will generally have a smaller scaling factor. It can be obtained by
solving of a transcendental equation \cite{Braaten-05}.

Several nuclear systems have been discussed as possible candidates for
Efimov states. The most promising system known so far is the $^{22}$C
halo nucleus which was found to display an extremely large matter
radius~\cite{Tanaka:2010} and is known to have a significant S-wave
component in the $n$-$^{20}$C system~\cite{Horiuchi:2006}.  See
Ref.~\cite{Acharya:2013aea} for a recent study of Efimov physics in
$^{22}$C.

Whether heavier two-neutron halos exist is still an open
question. Recently, there has been much interest, both experimentally
and theoretically, in determining precise values for masses,
understanding shell evolution and the location of the dripline in the
neutron rich calcium isotopes
\cite{Dilling,jdholt2012,hagen2012b,witek2013}. Coupled-cluster
calculations of neutron rich calcium isotopes that included coupling
to the scattering continuum and schematic three-nucleon forces,
suggested that there is an inversion of the $gds$ shell-model orbitals
in $^{53,55,61}$Ca. In particular it was suggested that a large
S-wave scattering length might occur in $^{61}$Ca with interesting
implications for $^{62}$Ca.

A conclusive statement on whether a halo is an Efimov state can
generally only be made if a sufficient number of observables is known
and if those fulfill relations dictated by universality. However,
typically only a very limited number of observables in these systems
is accessible experimentally. Recently, significant progress has been
made in microscopic calculations of low-energy nucleon-nucleus
scattering properties starting from realistic nucleon-nucleon
interactions \cite{RGM_NCSM, GFMC_scat,SCGF_O16,hagen2012c}. In this
Letter, we use the coupled-cluster method \cite{ccm} combined with
modern chiral effective theory interactions and follow the method
outlined in Ref.~\cite{hagen2012c} to compute the elastic scattering
of neutrons on $^{60}$Ca. We analyze the resulting phase shift data to
obtain quantitative estimates for the scattering length and the
effective range and show that a large scattering length can be
expected in this system.  The results obtained from ab inito
calculations are then used as input for the so-called halo effective
field theory (EFT) that describes the halo system in terms of its
effective degrees of freedom (core and valence nucleons)
\cite{Bertulani-02,BHvK2}. We use halo EFT to analyze the implications
of the coupled cluster results for the $^{60}$Ca-$n$-$n$
system. Specifically, we focus on the signals of Efimov physics that
are a consequence of the large scattering length in the $^{60}$Ca-$n$
and $n$-$n$ systems.

\paragraph{Hamiltonian and method -}
We perform coupled-cluster calculations for $^{60,61}$Ca starting
from the intrinsic $A-$nucleon Hamiltonian,
\begin{equation}
  \label{ham}
  \hat{H} = \sum_{1\le i<j\le A}\left({({\bf p}_i-{\bf p}_j)^2\over 2mA} + \hat{V}
    _{NN}^{(i,j)} + \hat{V}_{\rm 3N eff}^{(i,j)}\right).
\end{equation}
Here, the intrinsic kinetic energy depends on the mass number $A$.
The potential $\hat{V}_{NN}$ denotes the chiral $NN$ interaction at
next-to-next-to-next-to leading order~\cite{entem2003,machleidt2011}
(with cutoff $\Lambda=500$ MeV), and $\hat{V}_{\rm 3N eff}$ is a
schematic potential based on the in-medium chiral $NN$ interaction by
Holt~{\it et al.}~\cite{holt2009}. The potential $\hat{V}_{\rm 3N
  eff}$ results from integrating one nucleon in the leading-order
chiral three nucleon force (3NF) over the Fermi sphere with Fermi
momentum $k_F$ in symmetric nuclear matter.
\begin{figure}[t]
  \centerline{\includegraphics*[angle=0,clip=true,height=5cm]{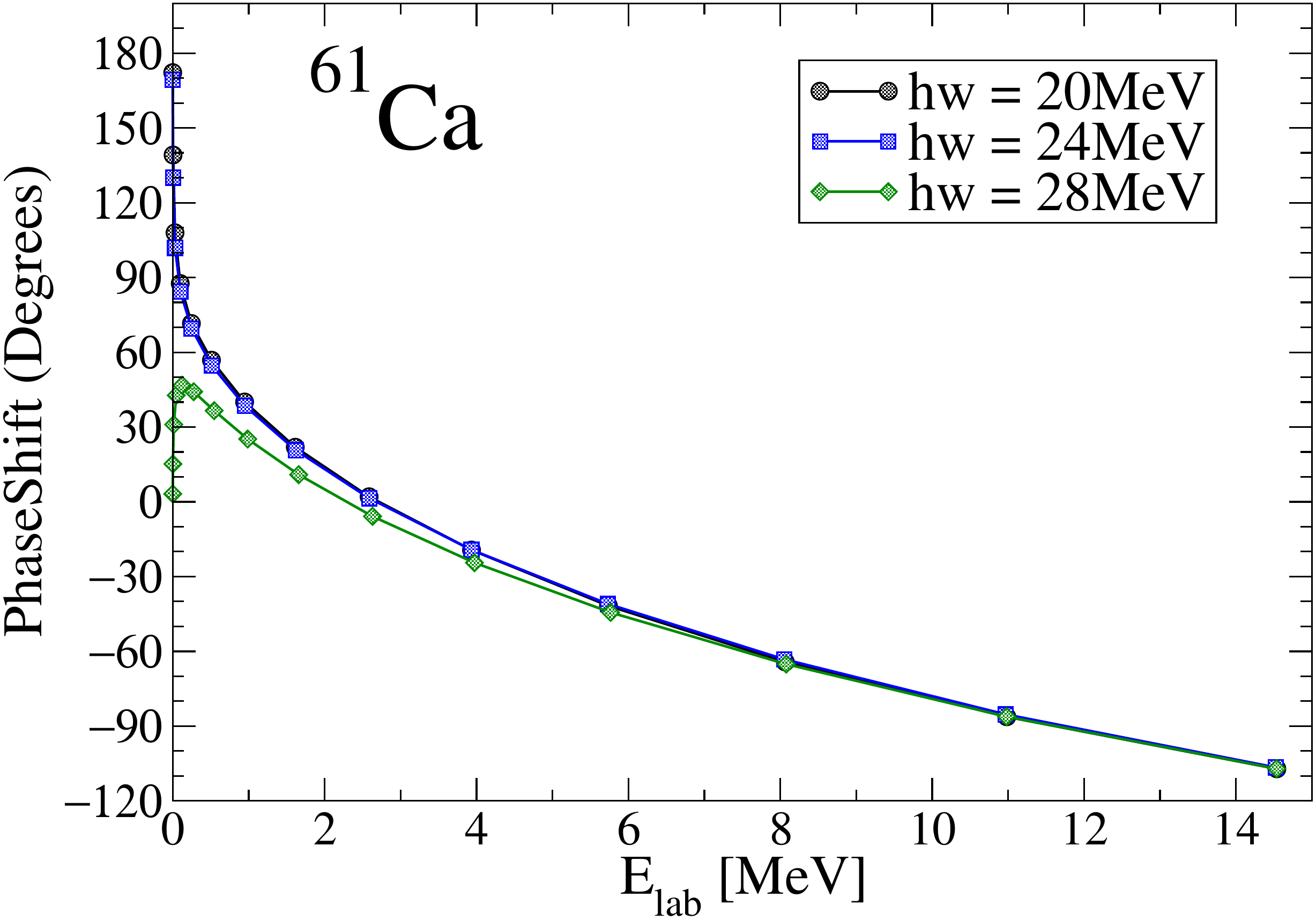}}
  \vspace*{0.0cm}
  \caption{(Color online) S-wave phase shifts for $n$-$^{60}$Ca scattering 
  from the coupled-cluster method as a function
   of the neutron energy in the lab frame for $\hbar\omega = 20,24,28$~MeV.}
  \label{fig:phaseshifts}
\end{figure}
In this work we employ the 3NF parameters which were used to study
shell evolution in neutron rich calcium isotopes \cite{hagen2012b},
and for proton elastic scattering on $^{40}$Ca \cite{hagen2012c}.
This interaction predicts well the masses for $^{50,51}$Ca
\cite{Dilling} and low-lying states in $^{53,54}$Ca
\cite{steppenbeck}, but is probably lacking in total binding energy
for isotopes around $^{60}$Ca.  To describe elastic scattering of a
nucleon on a nucleus $A$, we compute the one-nucleon overlap function
between the ground state of the nucleus $A$ with the scattering
solutions of the $A+1$ nucleus $O_{A}^{A+1}(lj;kr) =
\intsum_n\Braket{A+1||\tilde{a}_{nlj}^{\dagger}||A}\phi_{nlj}(r)$,
with the integral-sum over bound and scattering states (see
Ref.~\cite{hagen2012c} for more details). To obtain the ground state
of nucleus $A$, we use the coupled-cluster method in the
singles-and-doubles approximation (CCSD). For the excited states of
$A+1$, we use particle-attached equation-of-motion coupled-cluster
theory truncated at the two-particle-one-hole excitation level
\cite{bartlett2007}. We solve the coupled-cluster equations using a
Hartree-Fock basis built from $N_{\rm max}=17$ major spherical
oscillator shells and 50 Woods-Saxon discretized scattering states for
the S-wave \cite{hagen2012c}. We use the harmonic oscillator
frequencies $\hbar\omega = 20,24,28$~MeV to gauge the convergence of
our calculations.

We compute the radial overlap function $O_{A}^{A+1}(lj;kr)$ following
the method outlined in Ref.~\cite{jensen2010} and the elastic
scattering phase-shifts following the procedure described in
\cite{hagen2012c}. This amounts to matching the computed scattering
one-nucleon overlap functions to known asymptotic forms given by
spherical Bessel and Neumann functions. In Fig.~\ref{fig:phaseshifts},
we show the computed S-wave elastic scattering phase shifts for an
incoming neutron on the $0^+$ ground-state of $^{60}$Ca for different
harmonic oscillator frequencies. The corresponding CCSD ground state
energy for $^{60}$Ca shows a weak dependence on the harmonic
oscillator frequency, and we get $-386.07$~MeV,$-390.66$~MeV and
$-388.62$~MeV for $\hbar\omega = 20,24,28$~MeV, respectively. For
$\hbar\omega = 20$~MeV and $24$~MeV, $^{61}$Ca supports a very weakly
bound state, and the computed separation energy is $8$~keV and
$5$~keV, respectively. For $\hbar\omega = 28$~MeV, $^{61}$Ca is just
barely unbound.
In order to quantify the sensitivity of our results on the parameters
of our effective interaction, we varied the Fermi momentum $k_F$ in
our schematic 3NF by $\pm 0.01 ~\mathrm{fm}^{-1}$ away from the
optimal value $k_F = 0.95 ~\mathrm{fm}^{-1}$ \cite{hagen2012b}. We
found that the the total binding energy of $^{60}$Ca varied with $\pm
1$~MeV and the separation energy of $^{61}$Ca varied with $\pm
3$~keV. The uncertainty in our results coming from the tuned
parameters of our schematic 3NF is therefore of the same order as the
uncertainty coming from the finite size of the single-particle basis
used. Regarding the accuracy of our computed separation energy of
$^{61}$Ca we found that the $J^{\pi} = {1/2}^+$ state is dominated
(more than $95\%$) by one-particle excitations. This gives us
confidence that the $^{61}$Ca separation energy is much more
accurately computed than the total binding energy (which might be more
sensitive to correlations beyond the CCSD approximation). In
Ref.~\cite{hagen2012b} one of us computed the $J^{\pi} =
{5/2}^+,{9/2}^+$
excited states in $^{61}$Ca, and found them to be resonances with
energies $E_{{5/2}^+}=1.14-0.31i$~MeV and $E_{{9/2}^+}=2.19-0.01i$~MeV
with respect to the neutron emission threshold. Our results for S-wave
elastic scattering of neutrons on $^{60}$Ca therefore strongly support
the ground-state of $^{61}$Ca having $J^{\pi} = {1/2}^+$.

\paragraph{Scattering Parameters -}
The scattering phase shift data obtained
as described above, 
provides the input parameters required for a halo EFT analysis. For
low energies, the phase shifts for neutron-core scattering can be
represented by the effective range expansion
$  k \cot \delta_{cn}=-\frac{1}{a_{cn}}+\frac{r_{cn}}{2}k^2+\ldots~$,
where $k$ is the momentum in the center-of-mass frame and
the ellipses denote higher order terms in the expansion. We have
fitted the phase shift data
by
polynomials in $k^2$ and extracted the scattering length  $a_{cn}$ 
and effective range $r_{cn}$.  
The errors from the degree of the fitted polynomials are
negligible to the given accuracy. We obtained the scattering
parameters displayed in Tab.~\ref{tab:erepara}. 

\begin{table}[t]
\begin{tabular}{|c|c|c|c|c|}
\hline
$\hbar\omega$ [MeV] & $a_{cn}$ [fm] & $r_{cn}$ [fm] & $S_n$ [keV] &
$S_{\rm deep}$ [keV]\\
\hline
20 & 55.0 & 8.8&8.4  & 544 \\
24 & 53.2 & 9.1& 5.3 & 509\\
28 & -26.1 & 10.8& - & 361\\
\hline
\end{tabular}
\caption{Extracted $^{60}$Ca-$n$ scattering length $a_{cn}$ 
(1st column) and effective range $r_{cn}$ (2nd column) for different
oscillator parameters $\hbar\omega$. The neutron separation energy 
$S_n$ and the estimated breakdown scale $S_{\rm deep}$ are given in
the 3rd and 4th column, respectively.
\label{tab:erepara}}
\end{table}
For $\hbar\omega=28$ MeV, the scattering length is negative and the
$^{61}$Ca system is unbound. This could indicate that the implicit
infrared cutoff in the harmonic oscillator basis for $\hbar\omega=28$
MeV is too large to resolve threshold scattering.  While halo EFT
could
in principle 
be applied, there are fewer observables in this
case. Thus, we do not use these data in our analysis below. For
$\hbar\omega=20$ and $24$ MeV, we find consistent results.  The
scattering length is enhanced by about a factor of six compared to the
effective range. In the following, we will use the average of the
results for $\hbar\omega=20$ and $24$ MeV and take their spread as an optimistic
error estimate, i.e.
\beq a_{cn}=54(1)\ {\rm fm}
\qquad\mbox{and}\qquad r_{cn}=9.0(2)\ {\rm fm}~.
\label{eq:erevalues}
\eeq
The inverse effective range can be taken as an estimate of the
breakdown momentum beyond which the halo EFT cannot be applied
anymore.  The corresponding energy scale $S_{\rm deep}=1/(\mu_{cn}
r_{cn}^2)$ where $\mu_{cn}$ is the reduced mass of the $cn$
system is given in the 4th column of Table~\ref{tab:erepara}.

\paragraph{Halo EFT -} Halo EFT provides a model-independent
description of halo nuclei using the effective degrees of freedom of
these systems, i.e. the core and the valence nucleons.  Based on the
previous analysis, we will assume that the interaction is short-ranged
with $R\sim r_{cn}$. The expansion parameter of the EFT is $R$ divided
by the large scattering length $a_{cn}$.  The $^{60}$Ca-$n$
interaction is described by a spin-${\textstyle \frac{1}{2}}$ dimer
field $(\vec d_{cn})^T = (d_{cn,\uparrow}, d_{cn,\downarrow})$.  The
interaction of the two neutrons is described by a spin-$0$ dimer-field
$d_{nn}$. They have to be in the spin-singlet channel since they only
interact in the S-wave. To leading order in $R/a$, the Lagrangian is
then \cite{Bedaque:1998kg}
\begin{eqnarray}
\nonumber
     \mathcal L&=& \psi_c^\dagger \left(i\partial_0 + \frac{\nabla^2}{2M}\right)\psi_c
    +\vec \psi_n^\dagger \left(i\partial_0 + \frac{\nabla^2}{2m}\right) \vec \psi_n
    \\   \nonumber 
 &+& \Delta_{nn} \, d_{nn}^\dagger  d_{nn} \ + \Delta_{cn} \, \vec d_{cn}^{\;\dagger} 
  \vec d_{cn} +h\, d_{nn}^\dagger \psi_c^\dagger\, \psi_c\,d_{nn}\\  
&-&\left[g_{cn} \vec d_{cn}^{\;\dagger} \vec \psi_n \, \psi_c +  \frac{g_{nn}}{2}
  d_{nn}^\dagger \, (\vec \psi_n^\text{\;T} P \,\vec \psi_n) + \rm{h.c}\right]
  +\ldots~.
   \label{eq:lag_05}
\end{eqnarray}
where the $\psi_c$ and $\psi_n$ denote the core and neutron fields,
respectively, the ellipses denote higher-order terms and $P$ projects on the
spin singlet and the ellipses denote higher-order terms. The
coupling constants $g_i$ and $\Delta_i$ are fitted to the
effective range parameters of the $n$-$n$ and $^{60}$Ca-$n$ system.
Once this is done, various two-body observables can be calculated.
For example, the charge radius of a general one-neutron halo with
point-like core to next-to-leading order is~\cite{Hammer:2011ye}
\begin{eqnarray}
  \label{eq:1neutron-rc}
  \langle r_E^2 \rangle_{\rm rel}=\frac{{f}^2}{2\gamma_{cn}^2(1-r_{cn}\gamma_{cn})}~,
\end{eqnarray}
where $f=\mu_{cn}/M$, $M$ is the core mass, and $\gamma_{cn}$ the
binding momentum of the $cn$ system.  The total charge radius of
$^{61}$Ca is then obtained by adding this value to the charge radius
of $^{60}$Ca $\langle r_E^2\rangle_{{}^{61}\rm Ca}= \langle
r_E^2\rangle_{{}^{60}\rm Ca}+\langle r_E^2 \rangle_{\rm rel}$~.  Using
the scattering length and effective range from
Eq.~(\ref{eq:erevalues}), we obtain, $\langle r_E^2\rangle_{\rm rel} =
0.39(2)$~fm$^2$ where the error is from the uncertainty in $a_{cn}$
and higher order corrections are estimated to be of order
$(r_{cn}/a_{cn})^2\sim 3\%$. We note that the possible existence of a
low-lying excited state in $^{60}$Ca might introduce new parameters in
halo EFT and thereby alter the relations among $^{62}$Ca properties
discussed below.

Extending the framework of Ref.~\cite{Hammer:2011ye} to an external
current that couples to the matter distribution, we have calculated
the matter radius of a one-neutron halo. Normalizing the matter form
factor to unity, we find for the relative matter radius to next-to
leading order
\begin{eqnarray}
  \label{eq:1neutron-rm}
  \langle r_{mat}^2 \rangle_{\rm rel}=\frac{1}{2\gamma_{cn}^2(1-r_{cn}\gamma_{cn})}
\frac{\mu_{cn}}{M+m}~.
\end{eqnarray}
With the values from Eq.~(\ref{eq:erevalues}), we find $\langle
r_{mat}^2\rangle_{\rm rel} = 24(2)$~fm$^2$ where the error is
from higher order corrections which are
estimated to be of order $3\%$. Here, the matter radius of the core
should be comparable to $R$ and will
therefore give a sizeable but smaller contribution to the total
radius.

\paragraph{Three-Body Results -}
\begin{figure}[t]
\centerline{\includegraphics*[height=5cm, angle=0,clip=true]{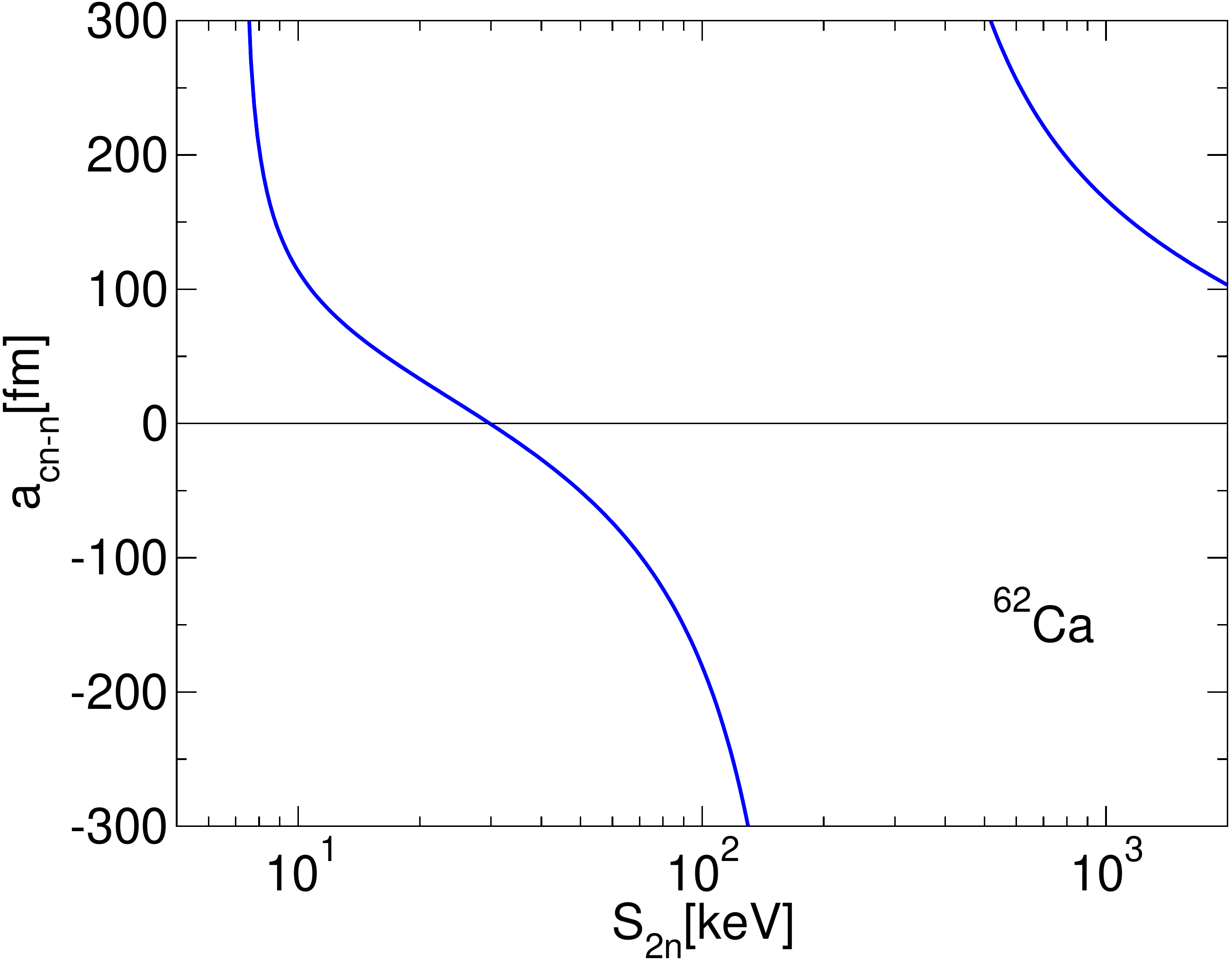}}
\vspace*{0.0cm}
\caption{The $n$-$^{61}$Ca scattering length as 
a function of the two-neutron separation energy $S_{2n}$.}
\label{fig:a_nd-B3}
\end{figure}
The large two-body scattering length
implies that the three-body sector ($cnn$) will display universal
features associated with Efimov physics. From the Lagrangian
(\ref{eq:lag_05}), we can derive a set of two coupled integral
equations for the $cnn$ system. Here, the three-body coupling $h$
contributes as well. (See Ref.~\cite{Hagen:2013xga} for more details.)
The bound state solutions of these equations exhibit discrete scale
invariance and the Efimov effect.  At leading order, the spectrum is
determined by the value of the scattering lengths in the two-body
sector and one observable in the three-body sector which is used to
fix the three-body coupling $h$ \cite{Bedaque:1998kg}.

For $^{62}$Ca, the discrete scaling factor governing the energy
spectrum is approximately $256$. The exact scaling symmetry
applies for deep states and in the unitary limit of infinite
scattering length. For two levels near threshold, however, the ratio
of their energies can be significantly smaller if one of the states is
very close to the threshold (see the discussion in~\cite{Braaten-05}
for the case of identical bosons). In our case, the whole energy
region between $S_n\approx 5-8$ keV and the breakdown scale $S_{\rm
  deep}\approx 500$ keV is available for Efimov states in
$^{62}$Ca. It is thus conceivable that $^{62}$Ca would display an
excited Efimov state and unlikely that it would not display any Efimov
states.

Another implication of the large scattering length in the
$cn$ system is that different low-energy three-body
observables are correlated. This means that the measurement of one
observable will uniquely determine all others up to corrections of
order of $R/a$. In Fig.~\ref{fig:a_nd-B3}, we display the correlation
between the two-neutron separation energy $S_{2n}$ of $^{62}$Ca and the
$n$-$^{61}$Ca scattering length $a_{cn-n}$. The scattering length
can take any value between $-\infty$ and $\infty$. When the binding
energy of the halo state $^{62}$Ca relative to the $n$-$^{61}$Ca
threshold vanishes,
the scattering length becomes infinite. The divergence in the
$n$-$^{61}$Ca scattering around 230 keV indicates therefore the
appearance of an additional state in the $^{62}$Ca spectrum.

\begin{figure}[t]
\centerline{
 \includegraphics*[height=5cm, angle=0,clip=true]{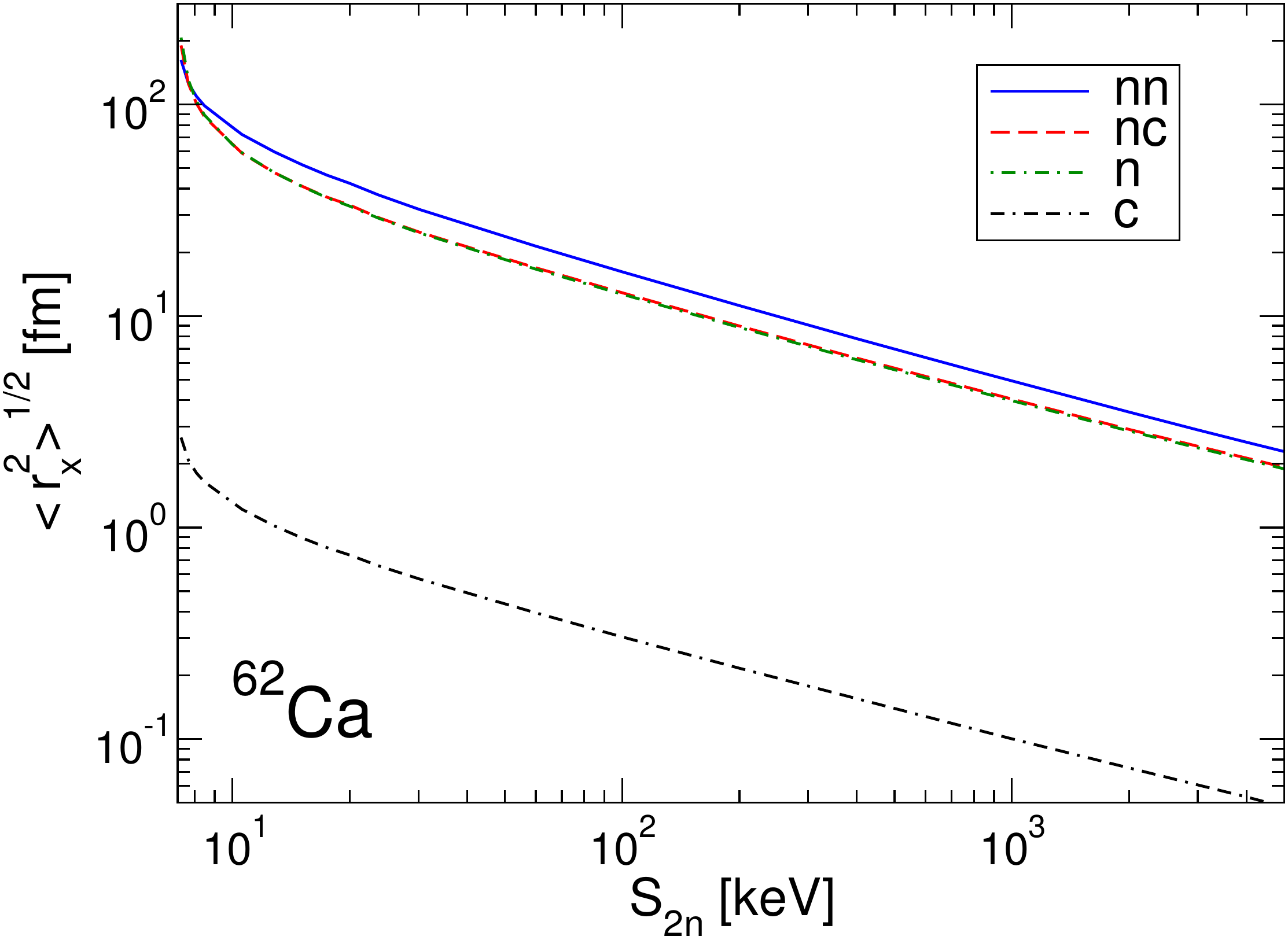}
}
\caption{\label{fig:matterradius} (Color online) Matter radii 
  $\langle r^2_x \rangle^{1/2}$ of the $^{62}$Ca system for $x=nn,nc,n,$
  and $c$
  as a function of the two-neutron separation energy $S_{2n}$}
\end{figure}
Four different matter radii can be calculated in the two-neutron halo
system. The mean square (ms) radius between the two neutrons ($\langle
r_{nn}^2\rangle$), the ms radius between the core and one of the
neutrons ($\langle r_{nc}^2\rangle$), and radii that give the ms
distance between the center-of-mass of the halo system and either one
of the neutrons ($\langle r_n^2\rangle$) or the core ($\langle
r_c^2\rangle$). These are correlated with other three-body
observables. We have calculated the various matter radii using the
methods of Ref.~\cite{Canham:2008jd} and show their correlation with
the two-neutron separation energy $S_{2n}$ in
Fig.~\ref{fig:matterradius}.

\begin{figure}[t]
\centerline{
\includegraphics*[height=5cm, angle=0,clip=true]{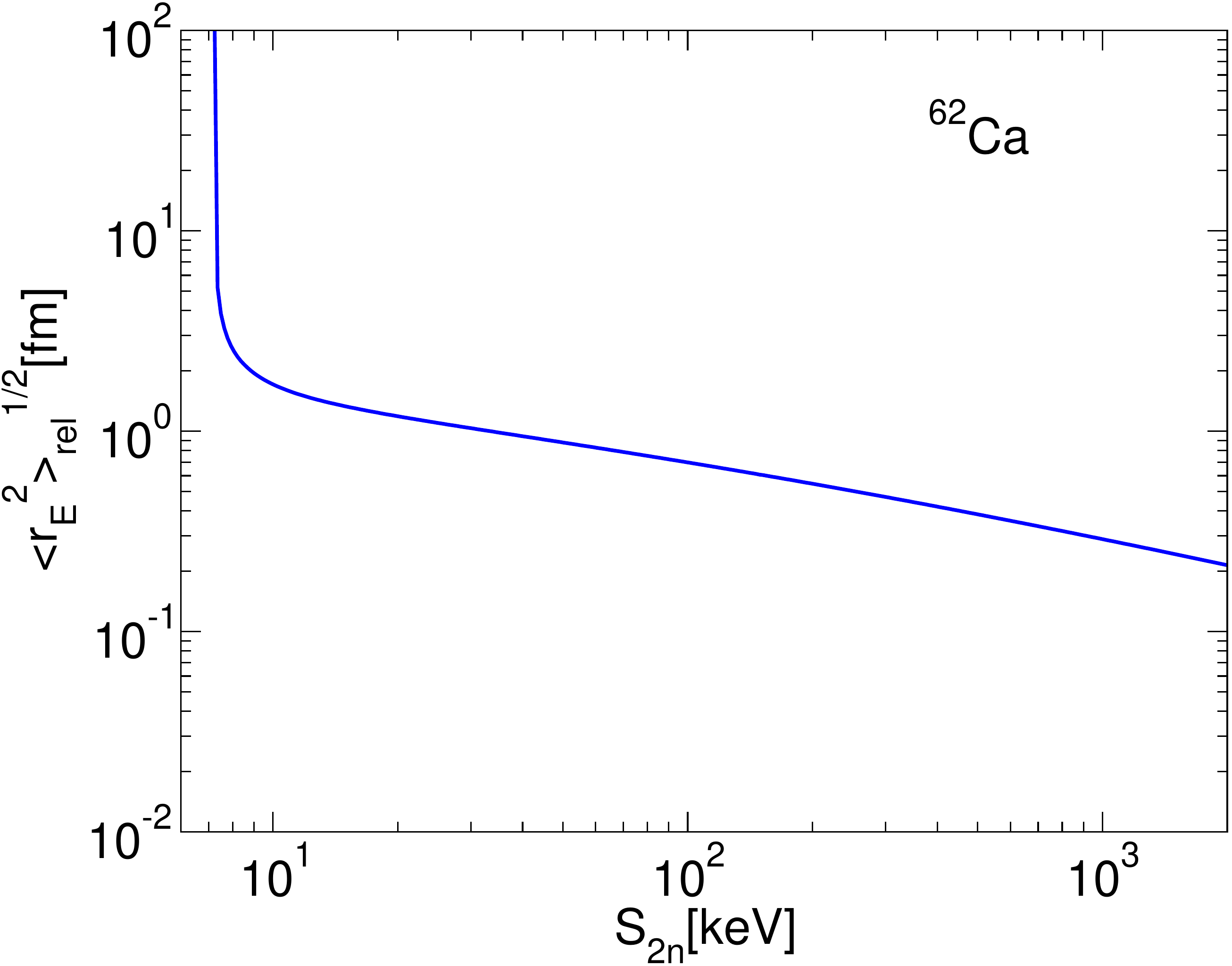}
}
\caption{\label{fig:chargeradius} (Color online) The relative charge radius 
  $\langle r^2_E \rangle_{\rm rel}^{1/2}$ of
  the $^{62}$Ca system as a function of the two-neutron separation
  energy $S_{2n}$.}
\end{figure}

The same type of correlation exists also for electromagnetic
low-energy observables. In Ref.~\cite{Hagen:2013xga}, the universal
correlation between charge radius relative to the core and the
two-neutron separation energy was studied as a function of the
core/nucleon mass ratio and the one- and two-neutron separation
energies of the halo nucleus. In Fig.~\ref{fig:chargeradius}, we show
this correlation for $^{62}$Ca. The total charge radius is expected to
be dominated by the $^{60}$Ca charge radius since the photon couples,
at leading order, only to the charged core and not to the neutrons. 

\paragraph{Summary -}
We have calculated the S-wave phase shifts of the $^{60}$Ca-$n$ system
using overlap functions obtained in coupled cluster theory. We
analyzed the phase shift data and combined the results with halo EFT
to predict universal features of the $^{61}$Ca and $^{62}$Ca
systems. Our analysis indicates, despite uncertainties in the coupled
cluster results due to truncation errors, a large scattering length in
the $^{60}$Ca-$n$ system. Specifically, the S-wave scattering length
is about 6 times larger than the effective range.  We calculated the
$^{61}$Ca-$n$ scattering length, and the matter and charge radii of
$^{62}$Ca. We have not considered the case of a negative scattering
length. In this case, the $^{62}$Ca system could be Borromean with
none of the two-body subsystems being bound. The features of the
$^{62}$Ca observables would qualitatively remain the same as those
depend mostly on the three-body parameter. From considerations based
on the scaling factor of this system and the breakdown scale of halo
EFT, we conclude that two Efimov states are possible in the $^{62}$Ca
system and that it is unlikely that this system possesses no bound
state, i.e. is unbound.

Our results imply that $^{62}$Ca is possibly the largest
and heaviest halo nucleus in the chart of nuclei. We have shown that
as a result a large number of observables would display characteristic
features that could be used to test our hypothesis. Measurements of
these observables will clearly pose a significant challenge for
experiment. For example, $^{58}$Ca is the heaviest Calcium isotope
that has been observed experimentally~\cite{tarasov2009}. However, the
planned FRIB might provide access to calcium isotopes as heavy as
$^{68}$Ca and thereby facilitate a test of our results
\cite{Sherrill:2012}.
\begin{acknowledgments}
  We thank T. Papenbrock for useful discussions.
  This work was supported by the Office of Nuclear Physics,
  U.S.~Department of Energy under Contract Nos. DE-AC02-06CH11357,
  DE-AC05-00OR22725 and DE-SC0008499 (NUCLEI SciDAC), by
  the DFG and the NSFC through funds provided to the Sino-German CRC
  110 ``Symmetries and the emergence of structure in QCD'', and by the
  BMBF under contract 05P12PDFTE.  Computer time was provided by the
  Innovative and Novel Computational Impact of Theory and Experiment
  (INCITE) program. This research used computational resources of the
  Oak Ridge Leadership Computing Facility and of the National Center
  for Computational Sciences, the National Institute for Computational
  Science.
\end{acknowledgments}

\end{document}